# Advancing consumer adoption of blockchain applications

Zane Witherspoon

*University of San Francisco*

**Abstract**

Blockchain technology as a whole is experiencing a dramatic rise in adoption, in no small part due to the developer-friendly Ethereum network. While the number of smart-contract powered distributed applications (Dapps) continues to rise, they face many of the same challenges all new technologies face as they are introduced to a market. By modeling the consumer adoption of blockchain technology and analyzing scholarly literature on supply-side factors affecting the diffusion of technology, we seek to prove the growth of a Dapp can be accelerated using *abstraction*, *whole product planning*, and *complimentaries*.

**1. Introduction**

Bitcoin has undisputedly been the most wildly successful application of blockchain technology. Created in 2008, the digital currency gained traction with the online drug market *The Silk Road* and the (now incarcerated) founder Ross Ulbricht, aka the Dread Pirate Roberts. Even after the FBI takedown of *The Silk Road* marketplace, the number of bitcoin wallet holders continued to climb at an exponential rate. At the start of 2014, there was still less than 1 million bitcoin wallets; by the start of 2017, there were almost 10 million[1].

Despite its incredible adoption of around the world, bitcoin is still far from market saturation. Bitcoin is still the most commonly used blockchain in the world, but other blockchains with their own unique functionalities and corresponding crypto-currencies are riding the wake of bitcoin's success to much faster adoption. Ethereum is the shining example of a new blockchain that's been adopted by even the most exclusive crypto-currency exchanges.

**Enter Ethereum**

The Ethereum organization took a developer-friendly approach when creating their own blockchain and introduced a new feature to the original bitcoin data storage technology: turing-complete programming languages for the Ethereum blockchain. Successful marketing and an ICO (Initial Coin Offering) attracted developers to build on the Ethereum blockchain faster than any other blockchain to date. This is credited in part to their coining of the phrase *smart-contracts* to define the code developers write to the blockchain and the celebritizing of Vitalik Buterin, the young Russian champion of Ethereum[2].

The challenges faced by Ethereum as a platform for distributed application (Dapp) development are those of a two-sided market. Imagine they are building a distributed app store. They need developers to build distributed apps for this new platform and consumers that are both able and willing to use this new tech. Ethereum is succeeding in the former, partly due to marketing and partly due to developers' tendencies to be technological innovators and visionaries. Targeting the latter part of this market is a much more challenging affair.

---

[1] *"Bitcoin Blockchain Wallet Users." Blockchain.info. Blockchain Luxembourg, n.d. Web. 10 May 2017.*
[2] Strang and Soule suggest that influence of an expert, or one percviced as so, accelerates social diffusion in organizations and social movements. *Strang, David, and Sarah A. Soule. "Diffusion in Organizations and Social Movements: From Hybrid Corn to Poison Pills." Annual Review of Sociology 24.1 (1998): 265-90. Web.*

**Consumer Adoption**

*"The rate at which new techniques are adopted and incorporated into the productive process is, without doubt, one of the central questions of economic growth."*
- *Nathan Rosenberg[3]*

The speed of the adoption of any new technology is one of the most challenging economic questions we face when looking into the rewards of innovation. Typically the innovations that have the greatest benefits are the same innovations that have the highest initial costs. The decision made by the consumer is not a matter of whether or not to adopt a new technology, but whether or not to defer the adoption to a later time when the initial costs have fallen[4].

The different ways of using blockchain technology have their own costs and benefits. As an example, using bitcoin as a currency has benefits like (mostly) anonymous money transfers, potential for financial profit traded against fiat currencies like the USD, and encrypted security so only you have access to your money. There are costs to using bitcoin as well, like the fiscal risk of a bitcoin price crash or the risk of losing an encryption key and access to your wallet. However given the right precautions, these costs are recoverable. As with other new technologies, the time spent learning the new technology is the most valuable, non-recoverable asset consumers have to pay.

Yet despite all these obstacles, companies like Coinbase and 21.co have had great success getting consumers to adopt bitcoin as a part of their products. While the ultimate decision to adopt a new technology is made by the demand side, these companies are successful because the benefits and costs of adopting a new technology can be affected by the supply side as well. In the end, it is an aggregation of decisions by both sides of the market that will determine the rate of diffusion[5].

---

[3] *Rosenberg, Nathan (1972). "Factors Affecting the Diffusion of Technology." Explorations in Economic History, Vol. 10(1), pp. 3-33. Reprinted in Rosenberg, N. (1976), Perspectives on Technology, Cambridge: Cambridge University Press, pp. 189-212.*
[4] *Hall, Bronwyn, and Beethika Khan. "Adoption of New Technology." National Bureau of Economic Research (2003): n. pag. Web.*

In this white paper, we will identify models that best represent the diffusion of blockchain technology on both sides of the market. By reviewing and applying the theoretical frameworks of scholarly literature to the current market drivers of blockchain technology, we'll uncover the apparent costs of adoption to the consumer. We will then explore the best practices by the suppliers of blockchain technology for mitigating those costs and elaborating the benefits as to increase the growth of their blockchain application.

**2. Review of Scholarly Literature**

The study of technological diffusion is well represented in academic literature. The study of blockchain technology is less so. For us to fully appreciate the fusion of these two subjects, usually studied by different industries entirely, we must accept that the proposals set forth, although novel, are built upon the foundation of academic study and industry research. We are working at the vertex of two very distinct areas of study to better understand what will accelerate the adoption of blockchain applications.

**Modeling the diffusion of technology**

Since we're discussing the diffusion of blockchain into the market, it is useful to visualize how that adoption looks over time. The most influential research on the diffusion of technology over time is <u>*Diffusion of Innovations*</u> by Everett Rogers[6]. Rogers argued that the diffusion of a new innovation looks like the S-curve in Figure 1. This curve visualizes the initial slow growth of a new technological innovation, followed by rapid growth by the market majority, and finally the slowing of growth as the market reaches saturation. The bell curve is the derivative of the S-curve and is frequently broken down into 5 major consumer categories: 1) Innovators 2) Early Adopters 3) Early Majority 4) Late Majority 5) Laggards.

---

[5] *Hall, Bronwyn, and Beethika Khan. "Adoption of New Technology." National Bureau of Economic Research (2003): n. pag. Web.*
[6] *Rogers, Everett M. Diffusion of Innovations. New York: Free, 2005. Print.*

The works of Rodgers are built upon by Geoffrey Moore in the book *Crossing the Chasm* [7]. Moore goes on to define the typical characteristics of the five categories of consumers. Moore uses these characterizations to group the categories by their expectations from the new technology. Moore argues that the key to winning the pragmatic majority is to plan for the **whole product** you offer. The whole product is the core product plus everything else you need to achieve your compelling reason to buy, including additional software, hardware, systems integration, installation and debugging, training and support, standards and procedures, etc.

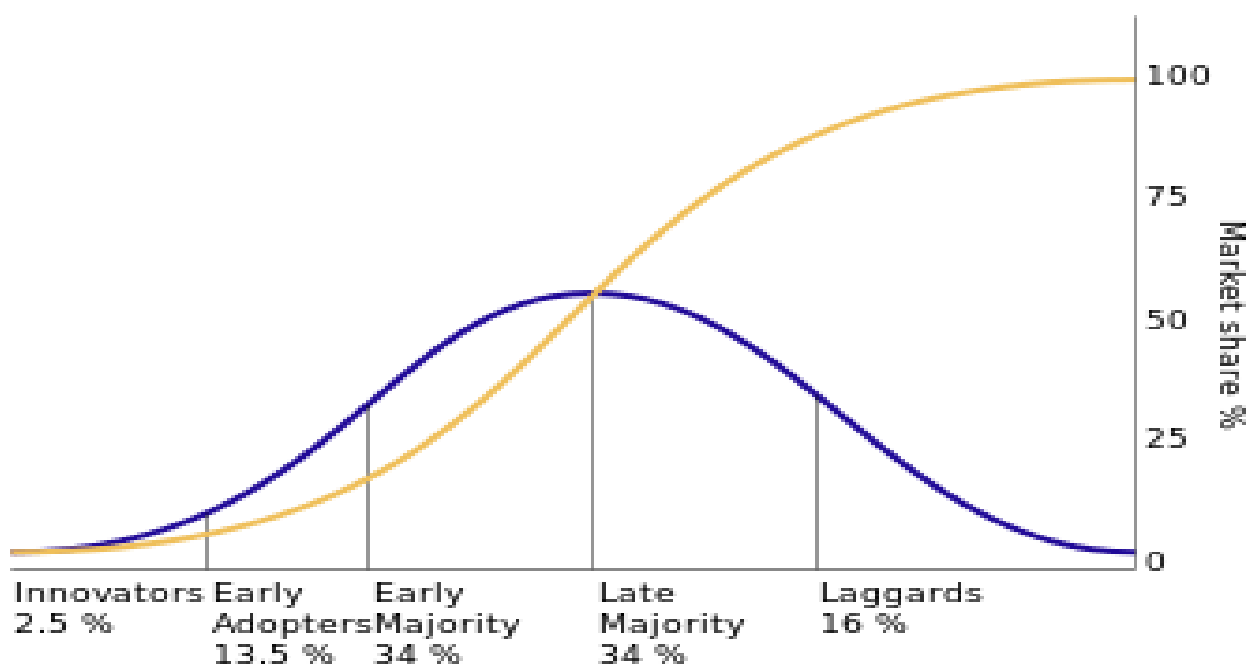

Figure 1

**Factors affecting the diffusion of technology**

For much of history, there was little research or even documentation on the diffusion of new technologies into a market. Some of the earliest and most influential work is credited to Nathan Rosenberg and his aptly named publication "Factors Affecting the Diffusion of Technology"[8] Here Rosenberg identified several of the reasons why the diffusion of a technology might be hindered or accelerated. While not all of these factors are applicable to blockchain (the development of skills in machine-manufacturing for example), there are some choice influences that cannot be denied as having an effect on the diffusion of blockchain technology into the market. First, Rosenberg puts forth the idea that continuous innovation of a new technology is what makes the second and third iterations of a new technology much more usable by the mainstream markets. Second, he proposes that an impediment in the diffusion of a new technology is the required *technical skill among users*. Lastly, *complementary* services and products that help the consumer to use the product accelerate the rate of diffusion of the technology. The idea of complementaries advancing the diffusion of a technology, supports Moore's whole product proposition as well.

---

[7] *Moore, Geoffrey A. Crossing the Chasm. Chichester: Capstone, (1998) Print.*
[8] *Rosenberg. "Factors Affecting the Diffusion of Technology." (1972)*

More recently, Bronwyn H. Hall and Beethika Khan of the University of California at Berkeley have built upon Nathan Rosenberg's findings and categorized them into three main groups: 1) Demand Determinants 2) Supply Behavior 3) Environmental and Institutional Factors[9]. Because this paper seeks to determine what behavior can lead to accelerated diffusion of a blockchain application, their research on supply behavior is particularly relevant. Namely, their elaborations on continuous innovation of new technology and complementary inputs.

**Blockchain as a technological innovation**

The first publication on blockchain technology was *"Bitcoin: A Peer-to-Peer Electronic Cash System"* published by the unknown creator of bitcoin under the pseudonym of Satoshi Nakamoto[10]. The white paper introduces some of the simplest blockchain applications like cash transfer and escrow, but doesn't begin to touch on the multitude of other applications for the new tech that other computer scientists have developed.

Most of the newest applications of blockchain technology have come to light after the introduction of the Ethereum network. One of the founders of Ethereum Dr. Gavin Wood explained how the Ethereum protocol can be used to build many different applications that work without a central database[11].

---

[9] Hall and Khan do a fantastic job of applying Nathan Rosenberg's work to modern innovations. The updated examples and classifications of the factors confirm Rosenberg's theories. *Hall, Bronwyn, and Beethika Khan. "Adoption of New Technology." (2003)*

[10] Nakamoto, Satoshi. "Bitcoin: A Peer-to-Peer Electronic Cash System." (2008): n. pag. Web.

> *"Each such project can be seen as a simple application on a decentralised, but singleton, compute resource. We can call this paradigm a transactional singleton machine with shared-state. Ethereum implements this paradigm in a generalised manner. Furthermore it provides a plurality of such resources, each with a distinct state and operating code but able to interact through a message-passing framework with others."*

Dr. Wood explains here the Ethereum Organization's emphasis on using the Ethereum blockchain as a platform to build projects. The Ethereum network went live in July 2015, and new applications are being built for it every day. It is important to note that this functionality is not exclusive to the Ethereum blockchain. Smart-contracts can in fact be written to the bitcoin blockchain and many others.

Most of the successful applications of blockchain so far are based in financial tech, but there are interesting projects being built around proof-of-existence (copyrights), physical asset-ownership (home or car titles), and gaming. Blockchain technology is still very much in it's infancy, and much like how the best smart-phone applications hadn't yet been considered after the release of the iPhone, I believe that the best blockchain applications have not yet been discovered.

## 3. Advancing Consumer Adoption of Your Blockchain Application

The costs and benefits of any particular blockchain application are as wide and varied as they are creative. For example a Dapp for picture storage would capitalize on the permanent qualities of the blockchain, but the transaction costs for storing large amounts of data would be astronomical. Storing your phone call history on the blockchain wouldn't be very costly, but the added benefits would be questionable. Our model of the diffusion of technology works under the assumptions that 1) a consumer that has made the transition to a new technology doesn't regress to the previous technology, and 2) the consumer will adopt the new technology once their added benefits outweigh their costs of adoption[12]. So while the cost and benefit variables of individual blockchain applications are different, the equations that govern their adoption are the same.

---

[11] Wood, Gavin. "Ethereum: A secure decentralised generalised transaction ledger." Ethereum Project Yellow Paper 151 (2014).

Knowing the factors affecting the diffusion of technology, we can identify some actionable practices by the suppliers of blockchain applications to facilitate more rapid adoption of their products. The next three sections of this article explain how **abstraction**, **whole product**, and **complementaries** can reduce costs or increase benefits of a blockchain application, therefore accelerating the diffusion.

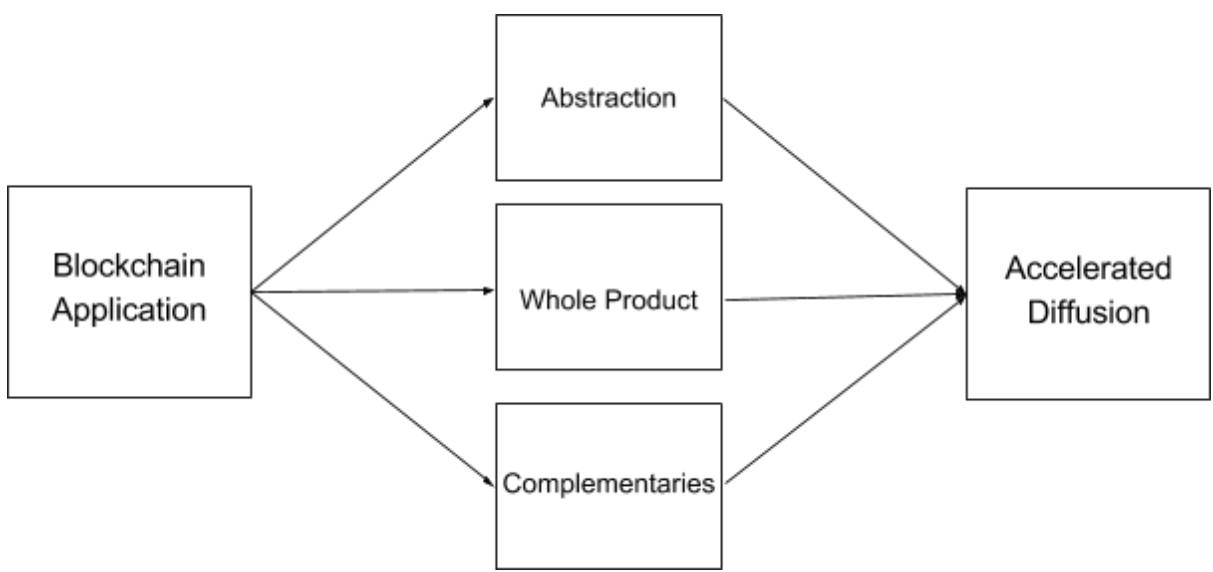

Figure 2

**Abstraction**

If you've ever filed your taxes online, you've already experienced some of the advantages of abstraction offers. By hiding the nasty IRS tax form behind a friendly, well-designed application, consumers can use the tech skills they already have to file their taxes. By taking the same approach to blockchain, we can enable a wider more consumers to use the tech skills they already have to use blockchain technology.

---

[12] Hall, Bronwyn, and Beethika Khan. "Adoption of New Technology." (2003)

Nathan Rosenberg proposes that one of the demand side determinants of technology adoption is the development of *technical skills among users*. There is, "a learning period the length of which will depend upon many factors, including the complexity of the new techniques, the extent to which they are novel or rely on skills already available."[13] Getting users to quickly use an application is therefore dependent on how novel or familiar the application experience is for the target consumer. Blockchain technology is a revolution in applied cryptography.

It has never been standard application practice to make a consumer hold onto their own encryption keys or use a currency from outside of their home countries, yet that's exactly what most distributed applications are asking of their customers today. That's not to say these applications are not successful, but that their costs of adoption are higher than traditional applications. Applying Rosenberg's principle of technical skills among users, it can easily be seen how requiring these novel skills from users will prolong the adoption of blockchain applications. It stands to reason then that by abstracting away these new skills currently required in a distributed application, the creators of blockchain applications can accelerate the adoption of their products.

**Whole Product**

Think about cable television as a product for a moment. The service is easy enough to purchase online or over the phone, but there's a gap between ordering the service and being able to enjoy your new subscription. The installation process is an ancillary service provided by cable providers to help fill that gap.

The concept of whole product as put forth by Geoffrey Moore is very straightforward: "there is a gap between the marketing promise made to the customer—the compelling value proposition—and the ability of the shipped product to fulfill that promise. For that gap to be overcome, the product must be augmented by a variety of services and ancillary products to become the whole product."[14] The whole product, therefore, serves to lower the costs of adoption for potential customers.

---

[13] *Rosenberg. "Factors Affecting the Diffusion of Technology." (1972)*
[14] *Moore, Geoffrey A. "Crossing the Chasm" (1988)*

The most basic smart-contracts on the Ethereum blockchain requires that the user run their own Ethereum node with their own local copy of the blockchain, learning how to interact with the blockchain using either a command line tool or the Ethereum wallet native application, and how to use the smart-contract itself. It's sounds like a simple call to action when asking a customer to try your smart-contract or distributed application by itself, but in reality, you're asking them to do much more. Asking customers to address these gaps in your product themselves is almost as costly to your growth as it would be for a cable provider to ask customers to install their own cable lines. Fill in the gaps of your product to accelerate your growth.

**Complementaries**

*"The importance of complementary inputs in the diffusion of new technology cannot be overemphasized"*
*- Hall & Kahn[15]*

Nathan Rosenberg explained complementaries by drawing from the history of American railroads in the second half of the nineteenth and the early twentieth centuries.[16] Hall and Khan explained complementaries by referencing makers of mobile telephones or PDAs teaming up with software suppliers like Microsoft to produce complementary software for customers to use. I'll continue this trend by examining successful complementaries developed by the blockchain company Coinbase.com.

Coinbase was originally and continues to be an exchange for converting fiat currency to bitcoin. This model turned out to be wildly successful, and competitors continued to enter the market. Coinbase has so far maintained their competitive advantage by offering complementary services and increasing the benefits of using their service. Customers who were interested in buying bitcoin were likely interested in buying other crypto-currencies as well, so in July 2016, 411Coinbase began trading the newer crypto-coin Ethereum.[17] In May 2017, Coinbase began trading Litecoin[18], and announced their plans to support even more crypto-currencies in the future. In addition to offering more currencies on their website, Coinbase also introduced GDAX.com, an exchange for serious traders to watch and place buy and sell orders to and from their favorite cryptocurrencies. By offering their customers other services they knew most would be interested in, Coinbase was able to increase the benefits of adopting their product, therefore, according to Hall and Kahn's principles of diffusion, accelerating their diffusion.

---

[15] Hall, Bronwyn, and Beethika Khan. "Adoption of New Technology." (2003)
[16] Rosenberg. "Factors Affecting the Diffusion of Technology." (1972)

## 4. Prepping for growth

Adoption of an application is a business problem with technological solution. Abstraction, whole product, and complementaries are all already being implemented by the best Dapps on the market. But much like the first companies to use the internet had no choice but set up their own servers, the first companies to implement these solutions had no choice but to build the infrastructure themselves. No company maintains their own web servers, thanks to cloud computing products like AWS, GCE, and Microsoft Azure. Blockchain infrastructure services are still blossoming but there are already third-party options for accelerating the development and adoption of your dapp.

**What to look for in a blockchain solution**

When looking for a blockchain infrastructure solution to accelerate the growth of your distributed application, many factors should be considered, including:

**Ease of Development**: The quicker you can build out your product, the quicker you can put it in your customer's hands. A good solution should be easy to build into your application with little special training.

**Compatibility Across Platforms:** Look for a solution that will support users across platforms. By supporting all the major browsers for web apps and both iOS and Android for mobile, you're keeping your potential market as wide as possible.

---

[17] *"Coinbase Adds Support for Ethereum – The Coinbase Blog." The Coinbase Blog. Coinbase, 21 July 2016. Web. 10 May 2017.*

[18] *Nandwani, Ankur. "Coinbase Adds Support for Litecoin – The Coinbase Blog." The Coinbase Blog. Coinbase, 03 May 2017. Web. 10 May 2017.*

**Seamless Integration:** You want your user experience to be as clean and easy as possible. Seek a solution that helps you abstract away the complexities of the blockchain for your end user.

**Optional Complementaries:** Increase the benefit of adopting your application. A good solution will provide accessory services that help enhance the fluidity and usability of your distributed app.

**Interpolated Data:** Don't be left in the dark about what who's using your smart-contract. Find a blockchain solution you can use to see how people are interacting with your smart-contract.

**HappyChain's Ethereum Infrastructure as a Service**

Designed to simplify smart-contract development, HappyChain is the easiest way to deploy smart-contracts to the Ethereum network. With instant deployment for common use-cases like tokens, there's no faster way to get a Dapp off the ground. HappyChain's REST-ful HTTP API works on any browser and on any mobile device. And because it connects to the Ethereum network over traditional HTTP, using a HappyChain powered application requires no additional technical skills from your users.

Companies building their own Ethereum infrastructure miss out on the economies-of-scale HappyChain offers. The entire blockchain ecosystem is developing quickly and so is HappyChain. The entire platform is growing to include a host of complementaries like smart-contract analytics, optional transaction encryption, and payment processing for your customers that don't have Ether yet. HappyChain is a comprehensive suite of smart-contract services will help your Dapp grow as effectively as possible.

**About HappyChain**

To find out more about HappyChain and its services, visit [https://happycha.in](https://happycha.in) or call (210)473-9870

**About the Author**

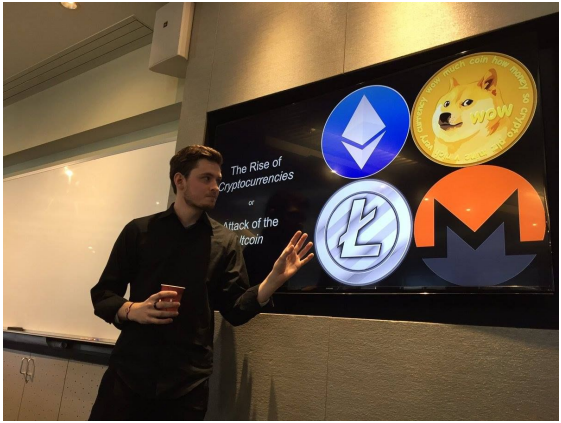

Zane Witherspoon is a founder of the blockchain infrastructure company HappyChain, organizer of the San Francisco Ethereum Developers Meetup, and a writer for the online publication HackerNoon. Witherspoon has a passion for writing and speaking about blockchain to audiences of all levels.